\documentstyle[aps,prd,epsfig]{revtex}
\newcommand{\bee}{\begin{equation}}
\newcommand{\ee}{\end{equation}}
\newcommand{\beea}{\begin{eqnarray}}
\newcommand{\eea}{\end{eqnarray}}

\preprint{COLO-HEP-471}
\begin{document}

\title{Another determination of
 the quark condensate from an overlap action}
\author{Thomas DeGrand}
\author{(MILC Collaboration)}
\address{
Department of Physics,
University of Colorado, 
        Boulder, CO 80309 USA
}
\date{\today}
\maketitle
\begin{abstract}
I use the technique of Hern\'andez, et al (hep-lat/0106011) to convert
a recent calculation of the lattice-regulated
 quark condensate from an overlap action
to a continuum-regulated number. I find
$\Sigma_{\overline{MS}}(\mu = 2$ GeV) = (282(6)   MeV$){}^3
\times (a^{-1}/1766$ MeV$){}^3$
from a calculation with the Wilson gauge action at $\beta=5.9$.
\end{abstract}
%
%

\section{Introduction}
Studying lattice QCD with fermion actions which respect chiral symmetry
without doubling  (via the Ginsparg-Wilson relation\cite{ref:GW})
is certainly a rewarding activity.
In the past two years there have been three calculations of the
(lattice-regulated)  quark condensate
$\Sigma$
in quenched QCD from overlap\cite{ref:neuberfer}
 actions \cite{ref:DEHN,ref:HJL,ref:TOM_OVER}.
Recently, Hern\'andez et al\cite{ref:HJLW}
 have adopted a matching of lattice and renormalization group invariant (RGI)
 masses to convert
their lattice regulated result for the quark condensate into its
 RGI counterpart.  The method involves determining the value of bare quark mass
at which the pseudoscalar mass takes on a certain value and
combining that bare mass with an appropriately-rescaled renormalization
factor computed using Wilson fermions \cite{ref:ALPUK},
 called $U_m$ by the authors of
Ref. \cite{ref:HJLW}. The $Z-$ factor which converts the lattice
 quark mass
to the RGI quark mass is then
\bee
\hat Z_M(g_0) = U_m {1\over{r_0 m_q}}|_{(r_0 m_{PS})^2=x_{ref}}.
\label{ZHAT}
\ee
Because Ginsparg-Wilson action are chiral, the renormalization
factor for the condensate is $Z_S= 1/Z_M$, and the RGI condensate
is
\bee
\hat \Sigma = {\Sigma \over {\hat Z_M}}.
\ee 
The RGI condensate can then be converted to the $\overline{MS}$ regulated
condensate using a table of (multi-loop) conversion coefficients
from Ref. \cite{ref:ALPUK}:
$\Sigma_{\overline{MS} }(\mu) = \hat \Sigma / z(\mu)$
where $z=0.72076$ for $\mu = 2$ GeV. My earlier calculation of $\Sigma$
\cite{ref:TOM_OVER} included a
heuristic calculation of the lattice-to-$\overline{MS}$ $Z-$
factor, which the work of Ref. \cite{ref:HJLW} renders obsolete.
It is easy to use  their results to update my old answer.
 The two fiducial points to which I can compare
are $x_{ref}=1.5376$ and 3.0, for which $U=0.181(6)$ and 0.349(9) respectively.

The data set used in this study consists of 40 $12^3\times 24$ lattices
generated
with the Wilson gauge action at $\beta=5.9$. The nominal lattice
spacing is $a\simeq 0.11$ fm,
using the Sommer radius $r_0=0.5$ fm
and the interpolating formula of Ref. \cite{ref:precis} of $r_0/a=4.483$.
Spectroscopy has been computed using the overlap action of
Ref. \cite{ref:TOM_OVER}, at bare quark mass $am_q=0.01$, 0.02, 0.04, 0.06.
This is an action whose ``kernel'' is an approximate overlap action, with
a connection to nearest and next-nearest neighbors and coupled to
the gauge fields through APE-blocked \cite{ref:APEblock} links.
Coulomb gauge Gaussian shell-model sources and point sinks were used.

\begin{table}
\begin{tabular}{|c|l|l|l|l}
\hline
$am_q$ & $r_0 m_q$ & $(r_0 m_{PS})^2$ &$(r_0 m_{PSS})^2$ & $(r_0 m_A)^2$ \\
\hline
   0.010 & 0.0448 &  0.853(118) & 0.332(82) & 1.088(203)  \\
   0.020 & 0.0897 &  0.968(138) & 0.946(86) & 1.175(92)  \\
   0.040 & 0.1793 &  1.928(144) & 2.164(82) & 2.022(78)  \\
   0.060 & 0.2690 &  2.505(165) & 3.067(126) & 2.949(80)  \\
\hline
\end{tabular}
\caption{Pseudoscalar masses (scaled by $r_0/a=4.483$) from the pseudoscalar
correlator ($m_{PS}$), the difference of pseudoscalar and scalar
 correlators ($m_{PSS}$), and from the axial vector correlator ($m_A$).}
\label{tab:pimass}
\end{table}

\begin{figure}[thb]
\epsfxsize=0.6 \hsize
\epsffile{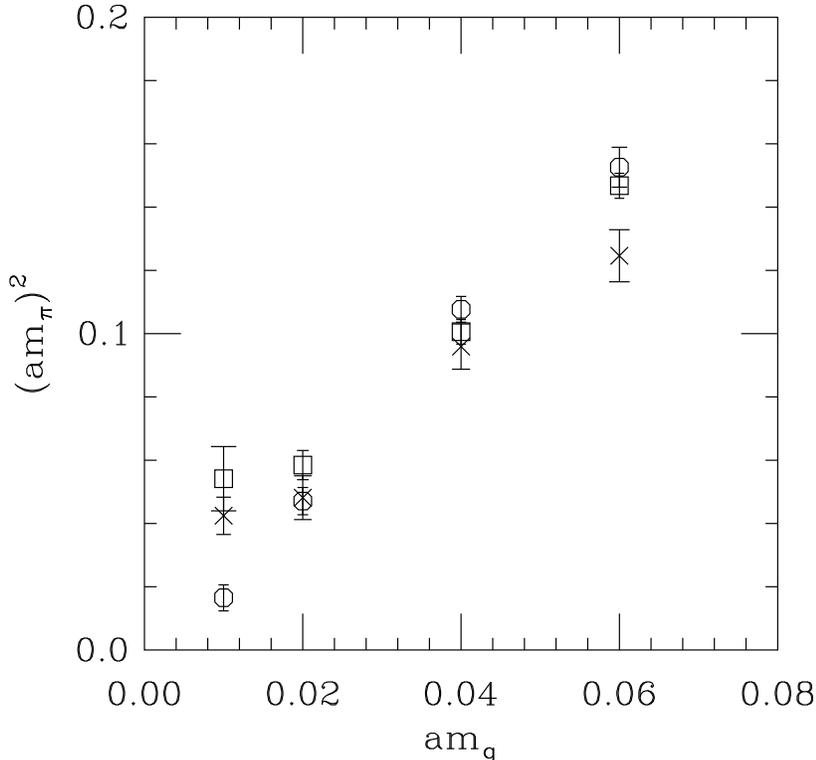}
\caption{
Squared pion mass vs. quark mass from the planar overlap. Crosses are masses
from the pseudoscalar correlator, octagons are from the difference of
pseudoscalar and scalar correlators, squares from the axial vector correlator.
}
\label{fig:pi}
\end{figure}

The data are displayed in Fig. \ref{fig:pi} and tabulated
 in Table \ref{tab:pimass}.

The overlap action has exact zero modes, which contribute a
quark mass independent finite-volume
lattice artifact in some channels, including the pseudoscalar (PS)
 and scalar channels\cite{ref:zeroes}.
 The difference of pseudoscalar and scalar channels
(which I will denote PSS below)
receives no contribution from zero modes in both quark propagators.
This is also the case for correlators of the axial current.
The effect of the zero modes is to flatten the meson correlator
and to give a meson mass in the PS channel which does not extrapolate to zero
at zero bare quark mass.
(In contrast, the PCAC quark mass extracted from this data set
does not show any additive mass renormalization, as expected.)
The PSS and axial channels do not suffer from this contamination.
As a result of this difficulty, I have elected to use the PSS
and axial data sets
to interpolate quark masses from meson masses.
At small quark masses the axial current decouples from the pion
(since $\langle 0 | \bar \psi \gamma_0 \gamma_5 \psi |\pi>= m_\pi f_\pi$)
and the signal in the channel degrades. Fortunately, the masses needed
for Eq. \ref{ZHAT} lie away from the smallest quark mass point.

The authors of Ref. \cite{ref:HJLW} did not see any problems with zero modes,
but their lightest quark mass is heavier than my two lightest ones.

I extracted quark masses at the fiducial pseudoscalar masses
via a combination of  single elimination jacknife
and $N-$point Lagrange interpolation, with $N=2$, 3, 4: in each
jacknife ensemble I fit the PSS and axial
 data sets, extracted pseudoscalar masses
and interpolated them to find each $r_0 m_q$. The uncertainty
on $r_0 m_q$ came from the jacknife. I found
$r_0 m_q= 0.130(3)$ and 0.267(8) at $x_{ref}=1.5376$ and 3.0, respectively.
These results combine with the tabulated $U_m$'s to give
 $\hat Z_M=1.392(56)$ and
1.307(52). I will assume that the difference between these two numbers
is a statistical effect and combine them,
continuing the analysis taking $\hat Z_M=1.34(6)$.
The uncertainty on this parameter is a bit smaller than
authors of Ref. \cite{ref:HJLW} found in their analysis; I believe that is
due to the fact that I can bracket the smaller quark mass point,
while in their case it is the smallest mass they measured.

The bare lattice regulated quark condensate from Ref. \cite{ref:TOM_OVER} is
 $\Sigma a^3 = 0.00394(16)$. Combining uncertainties in quadrature,
I have an RGI condensate of
\bee
a^3 \hat \Sigma = 0.00294(18)
\ee
or
\bee
r_0^3 \hat \Sigma = 0.265(16).
\ee
Ref. \cite{ref:HJLW} quotes 0.226(31) for the latter result, so we differ
by just over one standard deviation. Of course, this is a comparison of
two very different actions at one not-too-small lattice spacing.

Finally, taking the lattice spacing from $r_0$,
$\hat \Sigma = 0.0162(10)$ GeV${}^3$ or (253(5) MeV)${}^3
\times (a^{-1}/1766$ MeV$){}^3$ and 
$\Sigma_{\overline{MS}}(\mu = 2$ GeV) = (282(6)   MeV$){}^3
\times (a^{-1}/1766$ MeV$){}^3$.

I have not included any uncertainty in $r_0$ in this result.
The reader should be aware that using my extrapolation of the rho mass
to set the lattice spacing would result in a lattice spacing of $a=0.13$ fm
 rather than the result from the Sommer parameter of $a=0.11$ fm.

The lattice-to-$\overline {MS}$ $Z$-factor for this action at $\beta=5.9$
is $Z_M(\mu=2$ GeV) = $1.34(6) \times 0.72076=0.97(4)$. 
 In Ref. \cite{ref:TOM_OVER}
I had done a heuristic estimate of $Z_M\simeq 1.07$, based on experience with
perturbation theory for clover fermions with fat-link actions \cite{ref:BD}.
Unpublished comparisons of perturbative calculations and nonperturbative
measurements of  additive mass renormalization and vector
and axial current renormalization factors typically show values quite
close to unity for large values of APE smearing, with deviations in the 
two determinations of
$Z$'s of about 0.05 or so.

\section*{Acknowledgements}
I would like to thank Laurent Lellouch for his advice.
 This work was supported by the
U.~S. Department of Energy.


\end{document}